\documentstyle[12pt,amsfonts]{article}
\oddsidemargin--1mm
\begin{document}
\newcommand{\pl}{\partial}
\newcommand{\be}{\begin{equation}}
\newcommand{\ee}{\end{equation}}
\newcommand{\ba}{\begin{eqnarray}}
\newcommand{\ea}{\end{eqnarray}}
\newcommand{\ph}{PS_{ph}}
\newcommand{\mbf}[1]{\mbox{\boldmath$ #1$}}
\def\tint{{\textstyle\int}}
\def\bn{\begin{eqnarray}}     
\def\en{\end{eqnarray}}       
\def\<{\langle}
\def\>{\rangle}
\bibliographystyle{unsrt}

\begin{center}{\large\bf Towards a nonperturbative path integral 
in gauge theories}

\vskip 1cm
SERGEI V. SHABANOV \footnote{On leave from: {\em Laboratory of
Theoretical Physics, JINR, Dubna, Russia}} and JOHN R. KLAUDER 
\footnote{Also {\em Department of Physics}}

\vskip 0.5cm
{\em Department of Mathematics,
University of Florida, Gainesville, FL 32611, USA}

\end{center}

\begin{abstract}
We propose a modification of the Faddeev-Popov procedure to
construct a path integral representation for the transition
amplitude and the partition function for gauge theories whose
orbit space has a non-Euclidean geometry. Our approach is based
on the Kato-Trotter product formula modified appropriately to
incorporate the gauge invariance condition, and thereby 
equivalence to the Dirac operator formalism is
guaranteed by construction. The modified path
integral provides a solution to the Gribov obstruction 
as well as to the operator ordering problem when the orbit space
has curvature. A few explicit examples are given to illustrate
new features of the formalism developed. The method is applied to
the Kogut-Susskind lattice gauge theory to develop a nonperturbative
functional integral for a quantum Yang-Mills theory.   
Feynman's conjecture about a relation between the mass gap 
and the orbit space geometry in gluodynamics is discussed
in the framework of the modified path integral.
\end{abstract} 

{\bf 1. Motivations}. In what follows we consider only  gauge theories
of a special (Yang-Mills) type where gauge transformations are
linear transformations in the total configuration space. Let
${\cal X}$ be a configuration space which is assumed to be a 
Euclidean space ${\mathbb R}^N$ unless specified otherwise. 
It is a representation space of a gauge group ${\cal G}$ so 
the action of  ${\cal G}$ on ${\cal X}$ is given by a linear
transformation $x\rightarrow \Omega(\omega)x$. A formal
sum over paths for the action invariant under gauge transformations
would diverge. To regularize it, Faddeev and Popov have proposed
to insert the identity \cite{fp}
\begin{equation}
1 = \Delta_{FP}(x)\int d\mu(\Omega) \delta\left(\chi(\Omega x)\right)\ ,
\label{1}
\end{equation}
where $\chi(x)=0$ is a gauge condition and $\Delta_{FP}$ is the corresponding
Faddeev-Popov determinant, into the divergent path integral measure so
that after a change of variables $x\rightarrow \Omega x$ the (divergent) 
volume of the gauge group can be factorized out. The determinant
$\Delta_{FP}$ is calculated by doing 
the group averaging integral (\ref{1}).

This formalism has provided a solution to two important problems
in the {\em perturbative} path integral quantization of gauge 
theories: The unitarity of the perturbative S-matrix and 
the construction of a {\em local} effective gauge fixed action.
Later on it has been observed that the procedure becomes ill-defined
beyond perturbation theory \cite{gribov}. A geometrical
reason for the Gribov obstruction is that there is no gauge
condition which would provide a global parameterization of the 
gauge orbit space ${\cal M}={\cal X}/{\cal G}$ by a set
of affine coordinates without singularities \cite{singer}.
This situation is often rendered concrete in the vanishing
or even sign changing of the Faddeev-Popov determinant.
An additional problem expected in the nonperturbative 
regime is the operator ordering ambiguity \cite{christ}.
When the kinetic energy $\dot{x}^2$ is projected on 
the gauge orbit space ${\cal M}$, one gets $(\dot{h}, g_{ph}\dot{h})$,
where $h$ are local coordinates on ${\cal M}$ and $g_{ph}$ is
an induced metric on ${\cal M}$. The metric is not flat \cite{babelon}.
In order to establish a correspondence between the Dirac operator
formalism and the gauge-fixed path integral, the latter must be
appropriately defined to take into account the operator
ordering in the physical kinetic energy operator.
It is known that the formal path integral for motion on
a curved manifold is ambiguous \cite{dewitt}. 
Up to the
second order of perturbative
Yang-Mills theory, operator ordering corrections have no
effect on the physical S-matrix \cite{christ}. Their role
in a nonperturbative description is unknown. A standard argument
used in dimensional regularization, $i\hbar\delta^3(0) =0$,
(a commutator of canonically conjugated field variables
at the same spacetime point, 
which typically emerges in the operator ordering
terms) does not seem to be applicable because the validity of
dimensional regularization beyond perturbation theory 
is not justified \cite{zinn}. Finally, the space ${\cal M}$ may have a
nontrivial topology \cite{singer}. Due to the locality of
the Faddeev-Popov construction, global properties of ${\cal M}$
are lost in the path integral. On the other hand the topology
of the orbit space may be important for the spectrum of physical
excitations. Feynman conjectured that the mass gap in gluodynamics
in (2+1) spacetime might be caused by a compactification of
certain directions in the configuration space upon an identification     
of gauge equivalent configurations \cite{feynman}.

In this letter we propose a modification of the procedure (\ref{1})
which gives a solution to the above problems and leads to
a well-defined path integral on the orbit space ${\cal M}$.
Its validity is demonstrated with a few examples. 

\vskip 0.3cm
{\bf 2. A modified Kato-Trotter product formula}. Let $\hat{H}$
and $\hat{\sigma}_a$ be self-adjoint Hamiltonian and operators
of constraints, respectively. We also assume $[\hat{H},\hat{\sigma}_a]=0$
which holds for a sufficiently large class of physically interesting
gauge models. Let the gauge group be compact. Consider a projection operator
\be
\hat{\cal P}=\int d\mu(\omega) e^{i\omega^a\hat{\sigma}_a}\ ,\ \ \ \ 
\int d\mu(\omega) =1\ , \ \ \ \hat{\cal P}=\hat{\cal P}^\dagger=
\hat{\cal P}^2\ .
\label{2}
\ee
The operator (\ref{2}) projects the total Hilbert space onto
the Dirac subspace of gauge invariant states. If the gauge group
generated by $\hat{\sigma}_a$ is not compact, we take a sequence
$c_\delta\hat{\cal P}_\delta$ 
of rescaled projection operators $\hat{\cal P}_\delta$, where 
$\hat{\cal P}_\delta$ projects on the subspace $\sum \hat{\sigma}_a^2
\leq \delta^2$, and then consider the limit $\delta\rightarrow 0$
in a weak sense.
The coefficients $c_\delta$ can always be chosen so that the limit
exists and defines a Hilbert space isomorphic to the Dirac one \cite{kl97}.
Let the Hamiltonian have the form $\hat{H}=\hat{H}_0 + \hat{V}$,
where $V=V(x)$ is an everywhere regular 
potential and $\hat{H}_0= \hat{p}^2/2$
is a kinetic energy operator. The operators $\hat{H}_0$ and $\hat{V}$
are assumed to be gauge invariant, i.e., they commute with the constraint
operators\footnote{This assumption is justified for gauge theories of
the Yang-Mills type. More general cases will be considered elsewhere
because they are technically more involved and 
require a phase-space  
representation of quantum theory and a Wiener measure regularization
of the corresponding path integral as proposed in \cite{klsh1}.}.
Consider the evolution operator $\hat{U}_T = \exp(-iT\hat{H})$
in the total Hilbert space. The physical evolution operator is then
$\hat{U}_T^D = \hat{U}_T\hat{\cal P}$. Making use of the Kato-Trotter
product formula \cite{kt} for the evolution operator $\hat{U}_T$ we
get the following product formula for the gauge invariant evolution 
operator
\be
\hat{U}_T^D = \left(e^{-i\epsilon \hat{H}_0}e^{-i\epsilon \hat{V}}
\right)^n\hat{\cal P}= 
\left(e^{-i\epsilon \hat{H}_0}\hat{\cal P}e^{-i\epsilon \hat{V}}
\right)^n \equiv\left(\hat{U}_\epsilon^{0D}e^{-i\epsilon \hat{V}}\right)^n
\equiv \left(\hat{U}_\epsilon^D\right)^n\ ,
\label{3}
\ee
where $\epsilon = T/n$ and $n\rightarrow \infty$.  In our previous
work \cite{klsh1} we have shown that in the continuum limit 
the averaging gauge variables $\omega$ play the role of the Lagrange
multipliers for the constraints in the classical action. By making
use of the classical theory of Kolmogorov, we gave an explicit 
construction of a countably additive probability 
measure for the gauge functions
$\omega(t)$. One natural choice was such that {\em any} set of values of
$\omega(t)$ at  {\em any} set of distinct times is equally likely.
Now we show that this concept leads to a modification of the Faddeev-Popov
procedure. 

First we observe that by construction 
\be
U_T^D(x,\Omega x')=U_T^D(\Omega x, x')=U_T^D(x,x')
\label{4}
\ee
that is, the projected amplitude determines a genuine transition
amplitude on the orbit  space ${\cal M}$. To obtain its relation 
to the conventional gauge fixed amplitude, we introduce a set of
local coordinates $h$ on ${\cal M}$ via a change of variables
\be
x=\Omega (\omega) h \ ,
\label{5}
\ee
where the configurations $x=h$ satisfy a gauge condition $\chi(x)=0$,
i.e., $h\in {\cal H}\subset {\cal X}$ and without loss of generality
we may assume ${\cal H}\sim {\mathbb R}^M$ (unless specified otherwise) 
with $M$ being the number of physical degrees of freedom. 
The gauge invariant amplitude (\ref{4}) can be reduced 
on the gauge fixing surface ${\cal H}$:
$U_T^D(x,x')=U_T^D(h,h')$. This leads to a necessary modification
of the procedure (\ref{1}): Instead of averaging the gauge condition 
over the gauge group, we propose to average an infinitesimal evolution
operator kernel for a {\em free} motion as follows from the 
modified Kato-Trotter product formula (\ref{3})
\be
U_\epsilon^{0D}(h,h') =(2\pi i\epsilon)^{-N/2} \int d\mu(\omega)
e^{i(h-\Omega(\omega)h')^2/2\epsilon}\ .
\label{6}
\ee
Since $\epsilon\rightarrow 0$, the averaging integral can be done
to proper order in $\epsilon$
{\em explicitly} by the stationary phase approximation.
The kernel $U_\epsilon^D(h,h')$ is obtained by adding 
$-i\epsilon V(h)$ to the exponential in (\ref{6}) in accordance
with (\ref{3}) because $V(x)=V(h)$ thanks to the gauge invariance
of $V$.

Suppose there exist transformations of the new variables 
$\omega$ and $h$ in (\ref{5}), $\omega\rightarrow \omega +\omega_s$
and $h\rightarrow h_s$, such that $x(\omega, h)=x(\omega+\omega_s,h_s)$,
where all configurations $x_s=h_s$ also satisfy the gauge condition
$\chi(h)=\chi(h_s)=0$. From $\Omega(\omega +\omega_s)h_s=\Omega(\omega)h$
follows that $h_s=\Omega_s h$, i.e., $h_s$ is a Gribov copy of
$x=h$ on the gauge fixing surface $\chi(x)=0$. If ${\mathbb S}$ is a collection
of all residual gauge transformations, the modular domain $K$ is the quotient
${\cal H}/{\mathbb S}$. The relation (\ref{5}) determines
a one-to-one correspondence if $h\in K$ and $\omega$ parameterizes a 
group manifold. If $h,h'\in K$, then the following relation holds
\be
U_T^D(h,h_s')= U_T^D(h_s,h')=U_T^D(h,h')\ .
\label{7}
\ee
It determines an ${\mathbb S}-invariant$ continuation of 
the transition amplitude
to the {\em covering} space ${\cal H}$ of the modular
domain $K$. The Faddeev-Popov determinant specifies the volume
element on ${\cal M}$ \cite{babelon}, so 
for any integral involving only functions on ${\cal M}$,
$f(x)=f(\Omega x)$, one has
$\int dx f(x) = {\cal V}_G\int_K 
dh \Delta_{FP}(h) f(h)$, where ${\cal V}_G$ is the volume of the gauge group.   
Therefore the folding of two projected infinitesimal evolution
operators reads
\be
U_{2\epsilon}^{D}(h,h')= \int_K dh_1 \Delta_{FP}(h_1)
U_{\epsilon}^{D}(h,h_1)U_{\epsilon}^{D}(h_1,h')\ .
\label{8}
\ee
The restriction of the integration domain 
in (\ref{8}) does {\em not} lead to a formal restriction
of the integration domain in the path integral resulting 
from (\ref{3}) in the continuum limit as one might naively
expect from (\ref{8}). An important point is that, if $K\neq {\cal H}$ 
(the Gribov problem is present), then $U_\epsilon^{0D}(h,h')$ has 
{\em no} standard form of an infinitesimal free transition 
amplitude on a manifold proposed in \cite{dewitt} as we now
proceed to demonstrate.

The stationary phase approximation can be applied
before the reduction of $U_\epsilon^D(x,x')$ on a gauge
fixing surface. A {\em deviation} from the conventional
gauge-fixing procedure results from the fact that there
may be more than just one stationary point. 
We can always shift the origin of the averaging variable $\omega$ so that
one of the stationary points is at the origin $\omega=0$.
Let $\hat{\mbf{\delta}}_a$ be operators generating gauge transformations in
${\cal X}$. Decomposing the distance $(x-\Omega(\omega)x')^2$
in the vicinity of the stationary
point, we find $(x-x',\hat{\mbf{\delta}}_ax')=0$. So in the formal
continuum limit we get the condition 
$\sigma_a(\dot{x},x)\equiv 
(\dot{x},\hat{\mbf{\delta}}_ax)=0$ induced by the averaging procedure.
This is nothing but the Gauss law enforcement for trajectories 
contributing to $U_T^D(x,x')$. Suppose there exists a gauge condition
$\chi_a(x)=0$, which involves no time derivatives, such that a generic
configuration $x=h$ satisfying it also fulfills identically 
the Gauss law, i.e., $(\dot{h},\hat{\mbf{\delta}}_ah)\equiv 0$.
We will call it a {\em natural} gauge.
In this case all other stationary points in the integral (\ref{6})
are $\omega_c=\omega_s$ where $\Omega (\omega_s)=\Omega_s \in {\mathbb S}$.
Therefore we get a {\em sum} over the stationary points in the 
averaging integral (\ref{6}) if the Gribov problem is present.
Still, in the continuum limit we have to control all terms of
order $\epsilon$. This means that we need not only the leading
term in the stationary phase approximation of (\ref{6}) but also
the next two corrections to it. Therefore the group element $\Omega(\omega)$
should be decomposed up to order $\omega^4$ because $\omega^4/\epsilon
\sim \epsilon$ as one is easily convinced by rescaling the 
integration variable $\omega \rightarrow \sqrt{\epsilon}\,\omega$. 
The measure should also be decomposed up to the necessary order
to control the $\epsilon$-terms. The latter would yield 
{quantum corrections} to the classical potential associated
with the operator ordering in the kinetic energy operator
on the orbit space. We stress that the averaging procedure
gives a {\em unique} ordering so that the integral is invariant
under general coordinate transformations on ${\cal M}$, i.e.,
does not depend on the choice of $\chi$ (cf. (\ref{nlb})). Thus, 
\ba
U_\epsilon^{0D}(h,h')&=&(2\pi i\epsilon)^{-M/2}\sum_{\mathbb S}
D^{-1/2}(h,h_s')\left[e^{i(h-h_s')^2/2\epsilon - 
i\epsilon \bar{V}_q(h,h_s')} + O(\epsilon^2)\right]\ ,
\label{9}\\
&\equiv& \sum_{\mathbb S} D^{-1/2}(h,h_s') \tilde{U}_\epsilon(h,h_s')\ , 
\label{10}
\ea
where $D(h,h')$ is the conventional determinant arising in the
stationary phase approximation, 
and by $\bar{V}_q$ we denote a contribution
of all relevant corrections to the leading order.
The amplitude
$U_\epsilon^D(h,h')$ is obtained by adding $-i\epsilon V(h)$
to the exponential in (\ref{9}). Note that $V(h)=V(h_s)$ thanks
to the gauge invariance of the potential. 

In general, the equations $\sigma_a(\dot{x},x)=0$ are not
integrable therefore the functions $\chi_a$ discussed above
do not exist. In this case we consider two possibilities.
Let $\Omega_c(h,h')$ be the group element at a stationary
point in (\ref{6}). Decomposing the distance in the vicinity
of the stationary point we get $(h-\Omega_ch',\hat{\mbf{\delta}}_a
\Omega_ch')=0$. Let $\chi$ be such that the latter condition
is also satisfied if $h'$ is replaced by $h^\prime_s$. 
In this case the sum over the stationary points is again
given by (\ref{9}) where in the first term of the exponential
$h^\prime_s$ is replaced by $\Omega_c(h,h_s^\prime)h_s^\prime$.
In the most general case, the sum over stationary points does
not coincide with the sum over Gribov copies. However 
for sufficiently small $\epsilon$, the averaged
short-time transition amplitude can always be represented in
the form (\ref{10}). Note that as $\epsilon$ goes to zero,
$U_\epsilon^0(x,x')\rightarrow \delta^N(x-x')$. If we change
the variables in the delta function according to (\ref{5})
and then do the gauge group average, we get, in general,
a sum of delta functions on the gauge fixing surface.
From the symmetry of the change of variables (\ref{5})
it follows that the support of the averaged delta function
is formed by points $h=h^\prime_s$, $h'\in K$ and $h\in {\cal H}$, 
which means that 
the sum over ${\mathbb S}$ emerges again for sufficiently small $\epsilon$.
We will explain below how to calculate relevant corrections
to the leading order of the limit $\epsilon\rightarrow 0$. 

Let us replace 
$U_\epsilon^D(h,h_1)$ in (\ref{8}) by the sum (\ref{10}) and
make use of (\ref{7}) applied to the second kernel in (\ref{8}):
$U_\epsilon^D(h_1,h')=U_\epsilon^D(h_{1s},h')$. Since the measure
on the orbit space does not depend on a particular choice of
the modular domain, $dh_s\Delta_{FP}(h_s)=dh\Delta_{FP}(h)$,
we can extend the integration to the entire covering space
by removing the sum over ${\mathbb S}$
\be
U_{2\epsilon}^{D}(h,h')= \int dh_1 |\Delta_{FP}(h_1)|
D^{-1/2}(h,h_1)
\tilde{U}_{\epsilon}(h,h_1)U_{\epsilon}^{D}(h_1,h')\ .
\label{11}
\ee
The absolute value bars accounts for a possible sign change
of the Faddeev-Popov determinant on the gauge fixing surface.
The procedure can be repeated from left to right 
in the folding (\ref{3}), thus removing the restriction 
of the integration domain and the sum over copies in all
intermediate times $t\in (0,T)$. The sum over ${\mathbb S}$ for the 
initial configuration $h'$ {\em remains} in the integral.

Now we can formally take a continuum limit with the result
\be
U_T^D(h,h') = \sum_{\mathbb S}\left[
\Delta_{FP}(h)\Delta_{FP}(h_s')\right]^{-1/2}\!\!
\!\!\int\limits_{h(0)=h_s'}^{h(T)=h}\!\!\!{\cal D}h\sqrt{\det g_{ph}}
e^{i\int_0^Tdt[(\dot{h},g_{ph}\dot{h})/2 - V_q(h) -V(h)]} \ ,
\label{12}
\ee
where $g_{ph}$ is the induced metric on ${\cal M}$. The local 
density $\prod_t\sqrt{\det g_{ph}}$ should be understood 
as the result of the integration over the momenta in the 
corresponding time-sliced phase-space path integral where
the kinetic energy is $(p_h,g_{ph}^{-1}p_h)/2$ with $p_h$
being a canonical momentum for $h$. To obtain (\ref{12}),
one usually sets $h' = h -\Delta$ for $h=h(t)$ and $h'=h(t-\epsilon)$
in each intermediate moment of time $t$ and makes a decomposition
into the power series over $\Delta$. According to the relation,
obtained in \cite{babelon},
between the volume of a gauge orbit through $x=h$, the induced
metric $g_{ph}$, and the Faddeev-Popov determinant we
get $D(h'+\Delta,h')=\Delta_{FP}^2(h')/\det g_{ph}(h') + O(\Delta)$.
This relation explains the cancellation of the absolute value
of the Faddeev-Popov determinant in the folding (\ref{3})
made in accordance with the rule (\ref{11}). The term $\Delta^2/\epsilon$
in the exponential (\ref{9}) gives rise to the kinetic
energy $(\Delta,g_{ph}\Delta)/2\epsilon + O(\Delta^3)$. 

The most difficult part to calculate are the operator ordering
corrections $V_q(h)$ in the continuum limit. Here we remark that
$D(h,h')$ has to be decomposed up to order $\Delta^2$, while
the exponential in (\ref{9}) up to order $\Delta^4$ because 
the measure has support on paths for which $\Delta^2\sim \epsilon$
(i.e., $\Delta^4/\epsilon \sim \epsilon$). There is a technique,
called the equivalence rules for Lagrangian path integrals
on manifolds, which allows one to convert terms $\Delta^{2n}$
into terms $\epsilon^n$ and thereby to calculate
$V_q$ \cite{dewitt,lv82}
\be    
\Delta^{j_1}\cdots\Delta^{j_{2k}} \rightarrow (i\epsilon\hbar)^k
\sum_{p(j_1,\ldots , j_{2k})}
(g_{ph}^{-1})^{j_1j_2}\cdots(g_{ph}^{-1})^{j_{2k-1}j_{2k}}\ ,
\label{13}
\ee
where the sum is extended over all permutations of the 
indices $j$ to make the right-hand side of (\ref{13})
symmetric under permutations of the $j$'s. 
Following (\ref{13}) one derives the Schr\"odinger 
equation for the physical amlitude (\ref{12}).
The corresponding Hamiltonian operator on the orbit space has the form
\be
\hat{H}_{ph} = -1/2 \Delta_{FP}^{-1}\,\pl_j\left([g_{ph}^{-1}]^{jk}\Delta_{FP} 
\,\pl_k\right) + V(h)\ ,
\label{nlb}
\ee
where $\pl_j=\pl/\pl h_j$.
Observe that the kinetic energy in (\ref{nlb}) does {\em not}
coincide with the Laplace-Beltrami operator on ${\cal M}$
because $[\det g_{ph}]^{1/2}\neq \Delta_{FP}$ 
as shown in \cite{babelon}.
The operator (\ref{nlb}) is hermitian with respect to the scalar product
where the volume element is $dh \Delta_{FP}$. It is also invariant
under general coordinate transformations on ${\cal M}$, i.e., its
spectrum does {\em not} depend on the choice of local coordinates
on ${\cal M}$ and, therefore, is gauge invariant.   
We skip 
technical details of the derivation of 
(\ref{nlb}) because they are standard in the
path integral formalism and instead turn to examples to illustrate
the main features of the new projection formalism. But before
we do so, let us comment that the effects on the energy spectrum
caused by the modification of the path integral can be found
from the pole structure of the trace of  the resolvent
\ba
{\rm tr}\,\hat{R}(\tau) &=& {\rm tr}\,\left(\tau - i\hat{H}\right)^{-1}=
\int_0^\infty dT e^{-\tau T}\,{\rm tr}\,\hat{U}_T^D\ ,
\label{14}\\
{\rm tr}\,\hat{U}_T^D &=& \int_K dh\Delta_{FP}(h) U_T^D(h,h)\ .
\label{15}
\ea
{\bf 3. Examples}. Let us take the simplest example \cite{christ}
where $x\in {\mathbb R}^3$ and the gauge transformations are rotations
of $x$ about the origin. The orbit space is a space of concentric
spheres. The {\em natural} gauge is $x^i = \delta^{i1}h$.  
This gauge
is not complete because there are rotations that change the sign 
of $h$: $h\rightarrow \pm h$. These transformations form the group
${\mathbb S}={\mathbb Z}_2$. 
The modular domain is an open half-axis $K=(0,\infty)$.
The measure on the orbit space is obtained by introducing spherical
coordinates on ${\mathbb R}^3$ so that $h=|x|$, while the spherical angles
parameterize the gauge group manifold $SO(3)/SO(2)$. It is noteworthy
to observe that in the projection formalism there is no principal
difference between reducible and irreducible gauge theories. Here
we obviously have a reducible case because $x$ has a stationary
group $SO(2)$. The Faddeev-Popov determinant is defined with respect
to a subset of independent constraints. In this case, the constraints
are three components of the angular momentum. So we can take the 
angular momenta about the axes $i=2,3$ as independent constraints
and find that $\Delta_{FP}=h^2$ in the full accordance with the 
measure resulting from the spherical coordinates. The angular measure
must be normalized by $4\pi$ to fulfill the condition in (\ref{2}).
So Eq. (\ref{6}) assumes the form
\be
U_\epsilon^{0D}(h,h') =(2\pi i\epsilon)^{-3/2}\, 1/2
\int_0^\pi d\omega \sin\omega e^{i(h^2 +h'^2)/2\epsilon -
ihh'\cos\omega/\epsilon} \ .
\label{16}
\ee
Doing the averaging integral {\em exactly} and substituting
the result into (\ref{3}) we obtain
\be
U_T^D(h,h^\prime) = [hh^\prime]^{-1} 
\left\{U_T(h,h^\prime)-U_T(h,-h^\prime)\right\}\ ,
\label{17}
\ee
where the amplitude $U_T(h,h')$ is given by a standard path integral
for a one-dimensional system with the gauge-fixed 
action $S_{gf}[h]=\int_0^Tdt[\dot{h}^2/2
-V(h)]$. There is no operator ordering correction.

The stationary phase approximation leads to the {\em same} result
if one takes into account {\em all} the stationary points, i.e.,
$\omega_s=0$ and $\omega_s=\pi$.
The second stationary point is associated with the only
nontrivial residual gauge transformation $h'\rightarrow -h'$.
Here one also gets $\bar{V}_q=0$ (cf. (\ref{9})).
The same path integral can be derived from the Schr\"odinger
equation projected on the orbit space \cite{lvsh1}. It is not hard
to calculate the partition function (\ref{15}) for a harmonic oscillator
$V=x^2/2=h^2/2$ and find the resolvent (\ref{14}). The poles determine
the spectrum $E_n = 2n +3/2$. The distance between the energy levels
is 2, while the oscillator frequency in the Hamiltonian is 1. The
effect of the doubling would be lost, had we ignored the
redundant discrete gauge symmetry and calculated the partition
function by a standard path integral for the gauge-fixed
action $S_{gf}$.

A simple model
to illustrate the effect of the residual gauge symmetry
in systems with several physical degrees of freedom 
is a Yang-Mills theory in (1+0) spacetime \cite{lvsh1}. 
A total configuration space is a Lie algebra $X$, and the
gauge group acts in the adjoint representation $x\rightarrow \Omega
x\Omega^{-1}$. 
The quadratic form in the exponential (\ref{6}) is taken with respect
to the Killing form on $X$: $(x,x)
\equiv {\rm tr}\,\hat{x}^2$,  where $\hat{x}y\equiv [x,y]$
for any $x,y\in X$ and the commutator stands for 
the Lie bracket in $X$.
If a matrix representation is assumed,
for compact groups 
it can always be normalized to the ordinary matrix trace.
The Gauss law is $\sigma =[\dot{x},x]=0$,
therefore the natural gauge condition to parameterize ${\cal M}$  
is $x=h \in H$ where $H$ is a Cartan subalgebra (a maximal commutative
subalgebra of $X$). The system has $r=\dim H$ physical degrees of freedom.
The invariant
measure $d\mu(\omega)$ is proportional to $d\omega \det[(\exp\hat{\omega}-1)
\hat\omega^{-1}]$.
In the exponential in (\ref{6}) we get $(h-\exp(\hat{\omega})h')^2$.
An adjoint transformation
of $x$ to $H$  is not unique and determined modulo
the Weyl group ${\mathbb W}$ 
acting in $H$. The Weyl transformations are inequivalent 
compositions of reflections in the hyperplanes orthogonal to simple
roots in $H$ \cite{hel}. So ${\mathbb S}={\mathbb W}$ and 
${\cal M}\sim H/{\mathbb W}=K^+$, 
where $K^+$ is the Weyl chamber in $H$. 
The stationary points of the 
integral (\ref{6}) are $\omega_c =\omega_W$ where
the adjoint action of the group element $\exp\omega_W$ on $h$ transforms
the latter by a Weyl reflection.
Introducing the Cartan-Weyl basis \cite{hel}
one can find the Jacobian of the change of variables (\ref{5}) 
(the Faddeev-Popov determinant)  $\Delta_{FP}(h)=
\det\hat{h}=\kappa^2(h)$ where
$\kappa(h)=\prod_{\alpha>0}(\alpha,h)$; the product is extended over
all positive roots in $H$. The operator $\hat{h}$ acts 
in the orthogonal complement 
of the Cartan subalgebra. Note that the Cartan group is the stationary
group of $h\in H$, so $\omega$ should take its values in the
orthogonal supplement of the Cartan subalgebra.
The factor $D(h,h')$ is given by the determinant of the 
operator $\hat{h}\hat{h}'$.  
Doing the stationary phase approximation
in the averaging integral, and then taking the folding (\ref{3}) we
find
\be
U_T^D(h,h')= \sum_{\mathbb W} [\kappa(h)\kappa(h_W')]^{-1}U_T(h,h_W')\ .
\label{18}
\ee
Here the amplitude $U_T(h,h')$ is given by the conventional 
path integral
(no restriction of the integration domain to the Weyl chamber) 
with the action $\int dt(\dot{h}^2/2-V(h))$.
There is no operator ordering correction, and the physical metric
is Euclidean. The same integral follows from a direct solution of
the Schr\"odinger equation \cite{lvsh1}. The Faddeev-Popov determinant
vanishes at the hyperplanes $(h,\alpha)=0$. The boundary of the 
modular domain (the Weyl chamber) is formed by these hyperplanes 
when $\alpha$ runs over the simple roots. The amplitude (\ref{18})
remains {\em finite} at the singular points (the Gribov horizon) because
$\kappa(h_W)=p_W\kappa(h)$ where $p_W=\pm 1$ is the parity of the 
corresponding Weyl transformation (cf. also (\ref{17}): It remains
finite at $h=0$ or $h'=0$)). Contributions of paths hitting the 
horizon have a dramatic effect on the spectrum of the {\em isotropic}
oscillator $V=x^2/2=h^2/2$. Doing the path integral        
for $U_T(h,h')$ in (\ref{18}) and calculating the poles of the resolvent
(\ref{14}) we find \cite{lvsh1} $E_n= \nu_1n_1 +\cdots+\nu_rn_r +N/2$,
where $n_i$ are non-negative integers, $\nu_i$ are orders of invariant
irreducible symmetric tensors of the Lie algebra, and $r$ and $N$ 
are, respectively, the rank and dimension of the group
(e.g., for SU($r$+1), $\nu_i = i+1$, $i=1,2,...,r$). Thus, the physical
frequencies of the isotropic oscillator appear to be $\nu_i \neq 1$.
The partition function calculated by a
standard path integral for the gauge fixed action of the model
would lead to the spectrum of the isotropic oscillator (i.e., $\nu_i
\rightarrow 1$). 

To illustrate the effects of curvature of the orbit space, we
consider a simple gauge matrix model. Let $x$ be a real 2$\times$2
matrix subject to the gauge transformations $x\rightarrow
\Omega(\omega)x$ where $\Omega\in SO(2)$. An invariant
scalar product reads $(x,x')={\rm tr}\,x^Tx'$ with $x^T$ being
a transposed matrix $x$. The total configuration space is ${\mathbb R}^4$.  
Let $\mbf{\mbf{\varepsilon}}$ be a generator of SO(2), i.e., 
$\mbf{\mbf{\varepsilon}}_{ij}=
-\mbf{\mbf{\varepsilon}}_{ji}$, 
and $\mbf{\varepsilon}_{12}=1$. Then $\Omega(\omega)
=\exp(\omega\mbf{\varepsilon})$. 
The Gauss law enforced by the projection is $\sigma=
{\rm tr}\, \dot{x}^T\mbf{\varepsilon}x =0$. It is not integrable.
We parameterize ${\cal M}$ by triangular matrices $h_{21}\equiv 0$
(the gauge $x_{21}=0$).
 The residual gauge 
transformations form the group ${\mathbb S}={\mathbb Z}_2$: 
$h\rightarrow \pm h$.
The modular domain is a positive half-space
$h_{11}>0$. Calculating the Jacobian of the change of variables
(\ref{5}) for this model we find the Faddeev-Popov determinant
$\Delta_{FP}(h)=h_{11}$. The plane $h_{11}=0$ is the Gribov horizon.
The averaging measure in (\ref{6}) reads $(2\pi)^{-1}
d\omega$, where $\omega\in [0,2\pi )$. 
The quadratic form in the exponential in (\ref{6}) is
\be
\left(h-e^{\omega\mbf{\varepsilon}}h^\prime\right)^2 =
(h,h) +(h^{\prime },h^\prime)
-2(h,h')\cos\omega  -2(h,\mbf{\varepsilon} h^\prime)\sin\omega\ .
\label{19}
\ee
A distinguished feature of this model from those considered above
is that the stationary point is a function of $h$ and $h^\prime$.
Taking the derivative of (\ref{19}) with respect to $\omega$ and 
setting it to zero, we find 
\be
\omega_c  =\tan^{-1}\frac{(h,\mbf{\varepsilon} h^\prime)}
{(h,h^\prime)}\ ,\ \ \ \omega_c^s=\omega_c +\pi\ .
\ee
The second stationary point $\omega_c^s$ is associated with 
the Gribov transformation $h\rightarrow -h$. A geometrical
meaning of the transformation $h^\prime\rightarrow \exp(\omega_c
\mbf{\varepsilon})h^\prime$ is transparent. 
The distance $[(h-h^\prime)^2]^{1/2}$
between
two points on the gauge fixing plane  is greater
than the minimal distance between the two gauge orbits
through $x=h$ and $x^\prime =h^\prime$.
By shifting $x^\prime$ along the gauge orbit to 
 $x^\prime_c = \exp(\omega_c\mbf{\varepsilon})h^\prime$
a minimum of the distance between the orbits is achieved.
 In such a way the metric on the orbit
space emerges in the projection formalism. Its explicit form
can be found as has been explained in section 2. We substitute
$\omega =\omega_c(h,h')$ into (\ref{19}), set $h^\prime =
h-\Delta$ and decompose (\ref{19}) in a power series over
$\Delta$. The quadratic term (the leading term) determines
the metric. We get $(\Delta, g_{ph}(h)\Delta)=
(\Delta,\Delta) +(\Delta,\mbf{\varepsilon} h)
(\mbf{\varepsilon} h, \Delta)/(h,h)$. The metric is not flat.
The scalar Riemann curvature of ${\cal M}$ is $R=6/(h,h)=6/(x,x)$.
Note that it is gauge invariant as it should be because the 
curvature $R$ does not depend on the choice of local coordinates
on ${\cal M}$. In the stationary phase approximation 
the cosine and sine
in (\ref{19}) should be decomposed up to fourth order
in the vicinity of the stationary point. In this model 
quantum corrections do not vanish. The infinitesimal transition
amplitude on ${\cal M}$ assumes the form
\ba
U_\epsilon^D (h,h') &=& (2\pi i\epsilon)^{-3/2} \left[D^{-1/2}(h,h')
e^{iS_\epsilon(h,h')} + D^{-1/2}(h,-h') e^{iS_\epsilon(h,-h')}\right]\ ,\\
S_\epsilon(h,h')&=& [(h,h) +(h^\prime,h^\prime) -2D(h,h')]/(2\epsilon)
-\epsilon [8D(h,h')]^{-1} -\epsilon V(h)\ ,
\label{20}
\ea 
where $D(h,h') = (h,h^\prime)\cos\omega_c +
(h,\mbf{\varepsilon} h^\prime)\sin\omega_c$. Observe that
the middle term in (\ref{20}) is a quantum correction
to the classical potential in the short-time 
action (\ref{20}) (it is proportional to $\hbar^2$, if one
restores the Planck constant). The two first terms in (\ref{20}) 
determine ({\em uniquely\/}!) the operator ordering in the kinetic
energy operator on the orbit space in accordance with (\ref{nlb}). 

Finally, we consider the case of an infinite number of stationary
points in the averaging integral (\ref{6}). This would have an effect
that the modular domain $K$ may become compact, which, in turn, leads
to a {\em discrete} spectrum regardless of the details of the potential
and therefore gives a mechanism for the gap between the ground state
and the first excited state. A suitable soluble model is the Yang-Mills
theory in cylindrical space time \cite{rajeev,sem,sh93} where space
is compactified into a circle. A Yang-Mills theory without
matter in two dimensions has no physical degrees of freedom unless
the spacetime has a nontrivial topology. The latter is exactly our case.
The total configuration space ${\cal X}$ is the space of all periodic
connections $A(x+2\pi )=A(x)$
taking their values in the Lie algebra
of a compact Lie group. The gauge group ${\cal G}$ acts on ${\cal X}$
as $A\rightarrow {}^\Omega\! A=\Omega A\Omega^{-1} -i\Omega\pl\Omega^{-1}$, 
and $\Omega(x+2\pi)=\Omega(x)$
(i.e., the gauge group elements $\Omega$ can be continuously deformed
towards the group unit \cite{thooft,jackiw}). The exponential
in (\ref{6}) assumes the form $\int_0^{2\pi} dx{\rm tr}(A -{}^{\Omega}\! A')^2$
for any two configurations $A(x)$ and $A'(x)$. It is the 
distance between $A(x)$ and ${}^\Omega\! A^\prime(x)$ introduced by
Feynman \cite{feynman}. 
A rigorous definition of the averaging integral over ${\cal G}$ can be 
given within the Kogut-Susskind lattice gauge theory, to which 
we turn shortly. Note, for example, 
that $N$ in (\ref{6}) is infinite because the number of degrees of
freedom is determined by the number of Fourier modes of $A(x)$. 

Here we assume the existence of
a normalized averaging measure on ${\cal G}$. 
The gauge group average enforces the Gauss law 
$\sigma =\pl\dot{A}-i[A,\dot{A}]\equiv\nabla(A)\dot{A}=0$.
The orbit space ${\cal M}$
can be parameterized by constant connections $A(x) = h$ taking
their values in the Cartan subalgebra $H$ \cite{rajeev,sh93}. 
This is the natural gauge because $\sigma(\dot{h},h)\equiv 0$.
The number of the 
physical degrees of freedom equals the rank of the gauge group.
We set $\Omega(x)=\exp(i\omega(x))$.
The averaging function $\omega$ 
takes its values in the Lie algebra and its constant
(independent of $x$) Fourier mode does {\em not} have any component
in the Cartan subalgebra because the connection $A(x)=h$ is {\em invariant}
under constant gauge transformations from the Cartan subgroup.
The Gaussian integral over such functions can formally be done in (\ref{6})
yielding $D^{-1/2}(h,h')=\det{}^{-1/2}[-\nabla(h)\nabla(h')]$ which
is easy to compute in the Cartan-Weyl basis \cite{sh93}. 
In the lattice
formulation described below this procedure would not involve any infinite
factors in $D(h,h')$ as in the formal continuum limit. Here we draw attention
to the following feature. A projection of any connection $A(x)$ into
the subspace of constant Cartan connections $h$ is not unique and determined
modulo the affine Weyl transformations \cite{sh93}. The affine Weyl group 
${\mathbb W}_A$
is a semidirect product of the Weyl group 
${\mathbb W}$ and the group of translation
along the group unit lattice in the Cartan subalgebra \cite{hel}.
Thus, the number of stationary points in the averaging integral (\ref{6})
is {\em infinite} because the number of elements ${\mathbb W}_A$ is infinite:
$h\rightarrow h_s\equiv h_W + 2\alpha n_\alpha/(\alpha,\alpha)$,
where $h_W$ is a Weyl transformation of $h$, $\alpha$ any positive root,
$n_\alpha$ an integer. The modular
domain $K=H/{\mathbb W}_A$ 
is called the Weyl cell \cite{hel}. It is {\em compact}
in $H$. This has a significant effect on the spectrum of the system.
In the continuum limit the transition amplitude has the form (\ref{18})
where the sum is extended over the affine Weyl group (over an {\em infinite}
number of stationary points $h_s^\prime$)\cite{sh93}, 
the Faddeev-Popov determinant
reads $\Delta_{FP}(h) =\kappa^2(h)$ and $\kappa(h)=
\prod_{\alpha >0}\sin(h,\alpha)$. There is {\em no} potential
energy in $1+1$ dimensional Yang-Mills theory ($V(h)\equiv 0$ in (\ref{18})).
Thus, we get an amplitude for a {\em free} particle symmetrized over
the affine Weyl group. 

In the simplest SU(2) case, the system has
only one physical degree of freedom. The Weyl transformations are
reflections $h_W =\pm h$ and the translations over the group unit
lattice are shifts $h\rightarrow h + 2n$. The amplitude (\ref{18})
appears similar to the one for a particle in an infinite well or
on a circle.
The spectrum of such a system is {\em discrete}. In general,
we can calculate the poles of the resolvent (\ref{14}) for the 2D Yang-Mills
theory using our path integral formalism and find \cite{sh93}
the spectrum $E_n = E_0[(\gamma_n,\gamma_n) -(\rho,\rho)]$
where $E_0$ is a constant, 
$\rho$ is a half the sum of all positive roots, and $\gamma_n$
ranges over a lattice in the Weyl chamber that labels all irreducible
representation of the gauge group. 

Thus, an infinite number 
of stationary points in the averaging integral (\ref{6}) is, in fact,
evidence that the physical configuration space (or its 
subspace) is compact. This is the mathematical structure of the 
path integral which is to be sought when studying the relation 
between the mass
gap in gluodynamics and the compactness of the orbit space.    
Physically, the existence of an infinite number of the stationary
points separated by {\em different} distances (large and small)
implies that physical eigenstates of the Hamiltonian should exhibit
a {\em periodicity} in some directions in ${\cal X}$, which
may cause the mass gap as has been argued by Feynman \cite{feynman}.
 
\vskip 0.3cm     
{\bf 4. Kogut-Susskind lattice gauge theory}. To fulfill our 
program in realistic Yang-Mills theory, we need a regularization
at short distances to give a rigorous meaning to the averaging
integral (\ref{6}) which is a functional integral. The simplest
possibility is to set the system on a lattice. Since we still
want to have a Hamiltonian formalism, the time should remain
continuous. With such a choice of the regularization we
arrive at the Kogut-Susskind lattice gauge theory \cite{ks}.

Let $y$ be a three-vector whose components are integer-valued
in the lattice spacing unit $a$, i.e., $y$ labels the lattice
sites. With each link connecting the lattice sites $y$ and
$y+ia$, where $i$ is the unit vector in the direction of the
$i$th coordinate axis, we associate a group element $u_{y,i}$
called the link variable. The collection of all values of all link
variables forms the configuration space ${\cal X}$ of the system.
The gauge group ${\cal G}$ acts on ${\cal X}$ by the rule
$u_{y,i}\rightarrow \Omega_{y}u_{y,i}\Omega^{-1}_{y+ia}$.
Therefore the averaging measure is a product of the normalized
Haar measures at each site: $d\mu =\prod_yd\mu_H(\Omega_y)$.
The Kogut-Susskind Hamiltonian is $H=H_0 +V$ where
the potential $V$ is nothing but the color magnetic energy
on the lattice. The kinetic energy is the sum 
of kinetic energies of a rigid rotators associated with
each link variable
\be
H_0 = 1/2 \sum_{y,i}{\rm tr}\, (\dot{u}_{y,i}u^{-1}_{y,i})^2\ .
\ee
An important observation is that the coupling of the rotators
occurs only through the potential $V$. Therefore the "free"
amplitude $U_T^0(\{u\},\{u'\})$, where $\{u\}$ is a collection
of configurations of rotators $u_{y,i}$, is {\em factorized}
into a product of the amplitudes for each rotator $U_T^0(u_{y,i},
u_{y,i}^\prime)$. An explicit form of the latter is well-known 
\cite{marinov} (a transition amplitude of a particle moving
on a group manifold). This replaces the free motion 
amplitude in a Euclidean space in the averaging integral (\ref{6}).
The averaging integral 
\be
U_\epsilon^{0D}(\{u\},\{u^\prime\}) = 
\int \prod_y\left[d\mu_H(\Omega_{y})\right]
\prod_{y,i}U_\epsilon^0(u_{y,i},\Omega_yu^\prime_{y,i}\Omega_{y+ai})
\label{ks}
\ee
can be calculated via both the stationary
phase approximation and the decomposition
over the characters of the irreducible representations of the amplitudes
$U_T^0(u,u')$ proposed in \cite{marinov}. Thus, we have reduced
the problem of constructing the transition amplitude on the 
orbit space to a standard mathematical problem in group theory.      
It is important that the effects of the Riemann structure (metric) 
and topology of the orbit space are {\em completely} determined
by the averaging procedure proposed and do {\em not} depend on
the details of the self-interaction $V$. Note that the average in (\ref{ks})
induces a {\em nontrivial} coupling between the rotators because 
six link variables attached to the same site are transformed with 
the {\em same} group element. So the 
factorization of the amplitude on ${\cal M}$ is no longer possible. 
This is the result of the enforcement of the Gauss law
in the physical amplitude. 

Thus,  we get, in principle, 
a constructive and consistent
procedure to study the effects of the non-Euclidean geometry
of the orbit space on physical quantities like the spectrum of
low energy excitations.
In particular, in the strong coupling limit ($g$ large) the potential
$V$ can be treated by perturbation theory \cite{ks}.   
To introduce an explicit parameterization of the orbit space,
one can, for example, make use of Morse theory for 
the Kogut-Susskind lattice gauge theory as proposed in \cite{zw}.
In this case, the initial configuration $\{u^\prime\}$ in
the averaging integral (\ref{ks}) is taken
from the corresponding modular domain. The projected amplitude
has a natural gauge invariant continuation to the covering
space of the modular domain. 

\vskip 0.3cm
{\bf 5. Conclusions}.
We have proposed a self-consistent {\em path integral} quantization of
gauge theories which resolves the Gribov obstruction and provides
control for the operator ordering problem, and it does {\em not}  
rely explicitly on the Schr\"odinger equation on the orbit space.
This was our main goal. The continuum limit in the Kogut-Susskind lattice
gauge theory might still be a difficult analytical problem. However,
it is believed that the techniques we have developed might be useful to study 
the effects of the non-Euclidean geometry and topology of the orbit space 
in some physically reasonable approximations. We have also demonstrated
its effectiveness with several examples.

\end{document}